\documentclass[12pt]{article} 
 
\usepackage{scicite} 
 
\usepackage{times} 
 
\usepackage{graphics} 
 
\usepackage{isolatin1}      
 
\newenvironment{sciabstract}{%
\begin{quote} \bf} 
{\end{quote}}

\newcounter{lastnote} 
\newenvironment{scilastnote}{%
\setcounter{lastnote}{\value{enumiv}}%
\addtocounter{lastnote}{+1}%
\begin{list}%
{\arabic{lastnote}.} 
{\setlength{\leftmargin}{.22in}} 
{\setlength{\labelsep}{.5em}}} 
{\end{list}}

\title{\Huge \bf Large-scale, Decelerating, Relativistic X-ray Jets from the Microquasar \xte}

\author 
{S. Corbel$^{1}$, R.P. Fender$^{2}$, A.K. Tzioumis$^{3}$, J.A. Tomsick$^{4}$,  \\
J.A. Orosz$^{5}$, J.M. Miller$^{6}$, R. Wijnands$^{6}$, P. Kaaret$^{7}$ \\
\\
  \normalsize{$^{1}$Université Paris VII and Service
d'Astrophysique, CEA, CE-Saclay,}\\
  \normalsize{ 91191  Gif sur Yvette, France} \\
  \normalsize{$^{2}$Astronomical Institute `Anton Pannekoek',
University of Amsterdam, and Center for High}  \\ 
  \normalsize{Energy Astrophysics, Kruislaan 403, 1098 SJ Amsterdam, The
Netherlands}\\
  \normalsize{$^{3}$Australia Telescope National Facility,
CSIRO, P.O. Box 76, Epping NSW 1710, Australia} \\
  \normalsize{$^{4}$Center for Astrophysics and Space Sciences,
University of California  at San Diego, }  \\
  \normalsize{MS 0424, La Jolla, CA 92093, USA}\\
  \normalsize{$^{5}$Astronomical Institute, Utrecht University,
Postbus 80000, 3508 TA Utrecht, The Netherlands}\\
  \normalsize{$^{6}$Center for Space Research, MIT, NE80-6055, 77
Massachusetts Avenue, }\\
  \normalsize{Cambridge, MA 02139-4307, USA}\\
  \normalsize{$^{7}$Harvard-Smithsonian Center for Astrophysics, 60
Garden Street, Cambridge, MA 02138, USA}\\
}

\date{} 

\def\grs{GRS~1915$+$105}

\def\1e{1E~1740.7$-$2942}
\def\grsb{GRS~1758$-$258}

\def\xte{XTE~J1550$-$564}

\def\xtedixsept{XTE~J1748$-$288}  

\newcommand\arcdeg{\mbox{$^\circ$}}
\newcommand\degr{\arcdeg}
\def\arcmin{\hbox{$^\prime$}}
\def\arcsec{\hbox{$^{\prime\prime}$}}
\def\sm{\mbox{M$_\odot$}}

\begin{document}


\baselineskip24pt


\maketitle 


\begin{sciabstract}

We have discovered at x-ray and radio wavelengths 
large-scale moving jets from the microquasar \xte.  Plasma
ejected from near the black hole traveled at relativistic
velocities for at least four years.  We present
direct evidence for gradual deceleration in a relativistic
jet.  The broadband spectrum of the jets is consistent with
synchrotron emission from high energy (up to 10~TeV)
particles accelerated in shock waves formed within the
relativistic ejecta or by the interaction of the jets with
the interstellar medium.   \xte\ offers a unique opportunity
to study the dynamical evolution of relativistic jets on time
scales inaccessible for active galactic nuclei jets, with
implications for our understanding of relativistic jets from Galactic
x-ray binaries and active galactic nuclei.

\end{sciabstract}


Collimated relativistic jets produced by active galactic
nuclei \cite{bri84} (AGN) and by accretion-powered stellar
compact objects in sources called microquasars \cite{mir99}
are commonly observed at radio wavelengths.  Such jets are
produced close to black holes (supermassive ones in AGN and
stellar-mass ones in microquasars) and may help probe the
dynamics of matter being accreted in intense gravitational
fields.  The unprecedented sub-arc second resolution of the
{\em Chandra} x-ray observatory has recently allowed the
detection of x-ray jets in many AGNs.  Whereas the radio
emission of AGN jets is thought to originate from synchrotron
emission, the nature of the x-ray emission is still under
debate, but synchrotron or inverse Compton radiation
 are likely involved \cite{har02}.   Jets produced by Galactic black holes,
such as \xte\, should evolve much more rapidly than AGN jets
and, therefore, could provide insights to the dynamical
evolution of relativistic outflows and also to the processes of 
particle acceleration.  Here, we present the
first detection of large-scale, moving, relativistic jets
ejected from a Galactic black hole binary.

The x-ray transient \xte\ (Galactic longitude and latitude 
{\it l} = 325.88\degr, {\it b}~=~--1.83\degr) was discovered
by the All-Sky Monitor (ASM) aboard the {\em Rossi X-ray
Timing Explorer} (RXTE) on 7 September 1998 \cite{smi98}. 
Shortly after its discovery, a strong and brief (about one
day) x-ray flare was observed on 20 September 1998
\cite{sob00,hom01} and radio jets with apparent superluminal
velocities  ($>$ 2 c, where c is the speed of light) were
observed starting on  24 September 1998 \cite{han01}. 
Subsequent optical observations showed that the dynamical
mass of the compact object is $10.5 \pm 1.0\ \sm$, indicating
that the compact object in \xte\ is a black hole, revealed
the binary companion to be a low mass star, and led to a
distance estimate of about 5.3 kpc \cite{oro02}.

Following the re-appearance of x-ray emission from \xte\ in
early 2002 \cite{swa02}, we initiated a series of radio
observations with the Australia Telescope Compact Array
(ATCA).  Observations on 16 January 2002 showed renewed
activity at radio wavelengths from the \xte\ black hole
binary \cite{cor02}. These observations also revealed a new
radio source $\sim$ 22~arc sec to the west of the black hole
binary.  ATCA observations performed on 29 January 2002 (Fig.~1), 
in an array configuration allowing higher spatial
resolution imaging, showed that the western source had a
possible extension pointed toward \xte.  The position angle 
of this radio source relative
to \xte\ was --85.8\degr $\pm$ 1.0\arcdeg, which is
consistent  with the position angle (--86.1\degr $\pm$
0.8\arcdeg, \cite{han02}) of the western component of the
superluminal jet observed during the September 1998 radio
flare with long baseline interferometry \cite{han01}.

Prompted by the detection of the western source along the
axis of the jet from \xte, we searched archival data from
{\em Chandra} taken in 2000 for x-ray sources located along
the jet axis of \xte.  The field of view of \xte\ was imaged
by {\em Chandra} on 9 June, 21 August and 11 September 2000.
Examination of the 0.3--8 keV images (Fig.~2) revealed an
x-ray source $\sim$ 23 arc sec to the east of \xte\ at a
position angle of 93.8\degr $\pm$  0.9\degr\ from \xte, lying
along the axis of the eastern components of the radio
superluminal jets \cite{han01} (at a position angle of
$93.9\degr \pm 0.8\degr $; \cite{han02}).   The angular
separation between this eastern source and \xte\ increased
from $21.3 \pm 0.5$ arc sec on June 9 to $23.4 \pm 0.5$ arc
sec on September 11, implying that the eastern source moved
with an average proper motion of 21.2 $\pm$ 7.2 mas
day$^{-1}$ between these two observations.  This marks the
first time that an X-ray jet proper motion measurement has
been obtained for any accretion powered Galactic or
extra-galactic source.  Our radio observations (Fig.~1)
performed with ATCA between April 2000 and February 2001
showed a weak, decaying, and moving radio source consistent
with the position of the eastern x-ray source.  It was not
detected in February 2002 (Fig.~1) with a three sigma upper limit of
0.18 mJy at 3.5 cm.

With the discovery of the western radio source in early 2002,
we obtained a 30 ks {\em Chandra} observation on 11 March
2002. In the resulting 0.3--8 keV image (Fig.~2), three
sources were detected along the axis of the jet: the x-ray
binary \xte, an extended x-ray source at the position of the
western radio source, and a faint source that is $29.0 \pm
0.5$ arc sec east of \xte.  This weak x-ray source was the
eastern source that had smoothly decayed and moved by $5.7
\pm 0.7$ arc sec since September 2000.  The eastern source
was active during a period of at least two years (from April
2000 to March 2002).  

The most remarkable feature of this {\em Chandra} observation
is the discovery of x-ray emission associated with the
western radio source.  Both the radio and x-ray emission of
the western source appeared extended towards \xte, and the
morphology matched well between the two wavelengths.  Most
(70\%) of the x-ray emission was concentrated in the leading
peak which has a full width at half maximum (FWHM) of
1.2 arc sec.  A trailing tail, pointed back towards \xte,
gave a full width at 10\% of maximum intensity of 5 arc sec.

\vskip .2in      

The alignment of the eastern and western sources with the
axis of the jet observed on 24 September 1998 \cite{han01},
as well as the proper motion of the eastern source, imply
that both new sources are related to the jets of \xte. In
addition, both sources are likely connected with the
apparently superluminal ejecta from the brief and intense
flare of 20 September 1998 \cite{han01}.  Indeed, large scale
ejections of relativistic plasma (from \xte) have been
observed and resolved only during this occasion; radio
emission when detected at other epochs has been associated
with the compact jet of the low-hard x-ray spectral state
\cite{cor01}.  Also, the RXTE/ASM has not detected any other
x-ray flares similar to the large flare of 20 September 1998
in subsequent monitoring.  The fact that the eastern source
apparently moves faster (see below) than the western source
is consistent with the interpretation in which the eastern
source constitutes the jet that is pointing toward Earth (the
approaching jet) and the western source the receding jet.

With the positions of the eastern (and approaching) jet on 9
June 2000 and that of the western  (and receding) jet on 16
January 2002, we find average proper motions of 32.9 $\pm$
0.7 mas day$^{-1}$ and  18.3 $\pm$ 0.7 mas day$^{-1}$,
respectively. At a distance of 5.3 kpc \cite{oro02}, this
corresponds to average apparent velocities on the plane of
the sky of $1.0\ c$ and $0.6\ c$ for the eastern and western
jets, respectively.  The proper motion of 21.2 $\pm$ 7.2 mas
day$^{-1}$ measured by  {\em  Chandra} for the eastern jet
between 9 June 2000 and 11 September 2000 is significantly
smaller than  its corresponding average proper motion, which
indicates that the ejecta decelerated after the ejection.
This is confirmed by the {\em Chandra} detection of the
eastern jet in March 2002, implying an average proper motion
of 10.4 $\pm$ 0.9 mas day$^{-1}$ between 11 September 2000
and 11 March 2002. The relativistic plasma was originally
ejected at greater velocities, as the relative velocity was
initially greater than $2\ c$ (the initial proper motion was
greater than 57 mas day$^{-1}$ for the approaching jet,
\cite{han01}).  These observations provide the first direct
evidence for gradual deceleration of relativistic materials
in a jet.  Previous observations of other microquasars are
consistent with purely ballistic motions (e.g.\
\cite{mir99,fen99}) except for the system called \xtedixsept,
where, after ballistic ejections, the jet was observed to
stop suddenly, presumably following a collision with
environmental material \cite{kot00,rup02}.

The eastern and western jets have been detected up to angular
separations from \xte\ of 29~arc sec and 23~arc sec,
respectively, which correspond to projected physical
separations of 0.75~pc and 0.59~pc, respectively, for a
distance of 5.3~kpc \cite{oro02}. These are large distances
for moving relativistic ejecta (in \grs, the ejecta have been
observed to travel up to projected distance of 0.08 pc and on
a maximum time scale of four months, \cite{mir99}).
Persistent large scale (1-3 pc) jets have previously been
observed only at radio  wavelengths, e.g.\ \1e\ and \grsb\
\cite{mir92,rod92}, but without indication of associated
moving ejecta. Our observations reveal that the relativistic
ejecta of a Galactic black hole have been able to travel over
parsec scale distances at relativistic velocities during
several years.  An important aspect of our discovery is that
it provides the first direct evidence for a large-scale,
moving x-ray jet from any black hole (Galactic or in AGN). 

SS~433 is the only other x-ray binary for which large scale
(up to $\sim$ 40 arc min, i.e.\ several tens of parsecs),
non-thermal x-ray emission has been previously observed,
probably associated with  interactions of the jets with the
interstellar medium (ISM)
\cite{sew80,wat83,bri96,saf97,saf99}.  The helical pattern
observed in the lobes, at radio wavelengths, indicates a
connection between the lobes and the corkscrew pattern
associated with plasma ejection close to (on arc sec scale)
the core of the SS~433/W50 system \cite{dub98}.  However,
relativistic motion at large scales has not been observed in
SS~433.  We note that thermal X-ray emission arising from
moving relativistic ejecta, but only out to $\sim$~0.05~pc from the
compact object, has been reported in SS~433
\cite{mar02,mig02}.

Our results demonstrate that the emission from relativistic
ejecta of Galactic black holes can be observed at wavelengths
extending up to x-rays.  Future sensitive, high-resolution
observations of other Galactic black hole jets in the
infrared \cite{sam96}, optical, and x-rays bands may reveal
that broadband emission from relativistic ejecta of Galactic
black holes is more common than previously thought and offer
an exciting way to probe the physics of the jets.  AGN jets,
which were previously detected at radio and optical
wavelengths, are now known, with the advent of the {\em
Chandra} observatory, to often produce x-rays.  Whereas the
radio emission of AGN jets is thought to originate from
synchrotron emission, the nature of the x-ray emission has
not always been clearly identified. Although it is thought to
be non-thermal, it is not always known  whether synchrotron 
or inverse Compton radiation predominates for a particular object
\cite{har02,wil01,sam01,sam02}.  

The nature of the physical mechanism producing the emission
from the relativistic jets of \xte\ can be understood by
looking at the broadband spectrum, e.g.\ for the western jet
on 11 March 2002 (Fig. 3).  The position and morphology of
the radio and x-ray counterparts of the western jet are
consistent with each other (Figs 1 and 2) and the spectral 
energy distribution is consistent with a single power law (of
spectral index --0.660 $\pm$ 0.005).  These facts favor a
scenario in which the broadband emission from the jets is
synchrotron emission from high energy particles.  Similar
conclusions could be drawn for the eastern jet in 2000,
because the overall radio flux was also consistent with an
extrapolation  of the x-ray spectrum with a spectral index of
--0.6.  Detection of x-ray synchrotron emission would imply a
large Lorentz factor, of the order of 2 $\times$ 10$^7$
(corresponding to an energy of $\sim$ 10 TeV), for the x-ray
emitting electrons (under the equipartition assumption giving 
a magnetic field of $\sim$ 0.3 mG).

Acceleration in a shock wave is the most likely origin for
the very high energies required.  Shock waves could be
produced by internal instabilities \cite{har00} or by varying flow speeds within
the jet, as proposed to occur in some models of gamma-ray
bursts or AGNs \cite{ree94,kai00}.  If several relativistic
plasmoids were ejected around 24 September 1998 \cite{han01}
and their velocities were slightly different, then it would
have taken several months (maybe years) for them to collide. 
Such a collision would have produced shock waves, particle
acceleration, and the brightening of the jets.  

A more plausible alternative is that the shock waves are
produced when the jet material moving with bulk relativistic
speed interacts with the ISM (i.e. an external shock).  In
fact, the gradual deceleration we observed for the eastern
jet would be easily explained by such interactions.
Inhomogeneities in the ISM could also be at the origin of 
the brightening of the eastern and western jet at different
epochs.  The origin of the western jet is less clear as no
proper motion has been yet observed.  Future observations
will show whether or not the western jet is still moving and
together with high spatial resolution observations and
broadband spectra will be important in deciding between the
models (internal or external shocks).  Also, regular
observation of the jets of \xte\ would map the propagation of
the shocks and allow estimation of the energy dissipated in
the jets while decelerating in the ISM.  Therefore, \xte\
offers a unique opportunity to study the dynamical evolution
of relativistic jets on time scales inaccessible for AGN
jets, and has implications not only for the study of jets
from Galactic x-ray binaries, but also for our understanding
of relativistic jets from AGNs.

\vfill\eject


\bibliography{1550bib}

\bibliographystyle{Science}


\begin{scilastnote}

\item SC and JAT acknowledge useful conversations with A.
Celotti, S. Heinz and V. Dhawan.  PK acknowledges useful
discussions with H.\ Falcke and D.\ Harris.  SC thanks C.\
Bailyn, S.\ Chaty, D.\ Hannikainen and D.\ Hunstead for
providing information before publication and F.\ Mirabel for
a careful reading of this manuscript.  SC would like to thank
R.\ Ekers, D.\ McConnell, R. Norris, B.\ Sault and the ATCA
TAC for allowing the radio observations.  We thank H.\
Tananbaum for granting Director's Discretionary Time for the
{\it Chandra} observations and J.\ Nichols for rapid
processing of the data.  We have made use of observations
performed with ESO Melipal Telescope at the Paranal
Observatory under Director's Discretionary Time programme
268.D-5771.  The Australia Telescope is funded by the
Commonwealth of Australia for operation as a National
Facility managed by CSIRO. 

\end{scilastnote}

\clearpage

\begin{figure*}[p] {\par\centering
\resizebox*{1\textwidth}{!}{
\rotatebox{-0}{\includegraphics{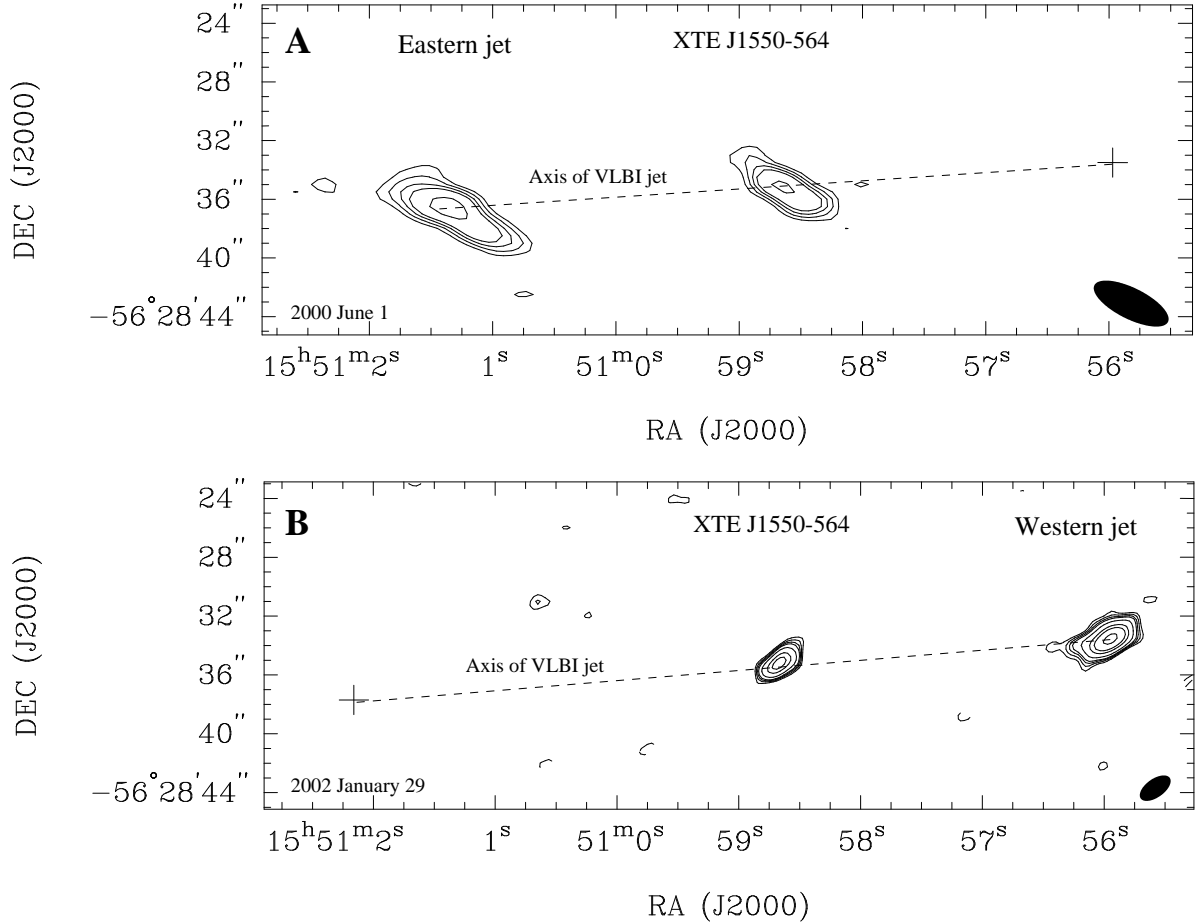}}} \par}
\caption{Uniform weighted maps of the field of view around
the black hole candidate \xte\ on 1 June 2000 (top) and 29
January 2002 (bottom) showing the radio counterpart to the
eastern and western jets (when detected). The stationary
black hole binary \xte\ is at the center of the image, and
has a radio spectrum typical of the self-absorbed compact jet
\cite{hje88,fen01} that is observed during the x-ray low/hard
spectral state (ref 12).  ({\bf A}) (1 June 2000): Map at
4800 MHz (6 cm). The position of the eastern radio jet is
$\alpha$(J2000) = 15h 51m 01s.30 and $\delta$(J2000) =
--56\degr\ 28\arcmin\ 36.9\arcsec\  with a total uncertainty
of 0.3 arc sec, i.e. at a position angle of  93.8\degr $\pm$ 
0.9\degr\ from \xte.  The synthesized beam (in the lower right
corner) is  $5.5\times 2.1$ arc sec with the major axis in
position angle of 63.1\degr. The peak brightness in the 
image is 1.1 mJy per beam. Contours are plotted at --3, 3, 4,
5, 6, 9 times the r.m.s. noise equal to 0.1 mJy per beam. The
cross marks the position of  the western jet, as measured on
29 January 2002. ({\bf B}) (29 January 2002):  Map at 8640
MHz (3.5 cm). The position of the western radio jet is
$\alpha$(J2000) = 15h 50m 55s.94 and $\delta$(J2000) =
--56\degr\ 28\arcmin\ 33.5\arcsec\ with a total uncertainty
of 0.3 arc sec. The synthesized beam is $2.4 \times 1.3$ arc
sec with the major axis in position angle of --54.6\degr. The
peak brightness in the  image is 1.79 mJy per beam. Contours
are plotted at --3, 3, 4, 5, 6, 9, 15, 20 30 times the r.m.s.
noise equal to 0.05 mJy per beam. The cross marks the
position of the eastern jet, as measured on 11 March 2002
during  the Chandra observation.} \end{figure*}

\begin{figure*}[p] {\par\centering
\resizebox*{1\textwidth}{!}{
\rotatebox{0}{\includegraphics{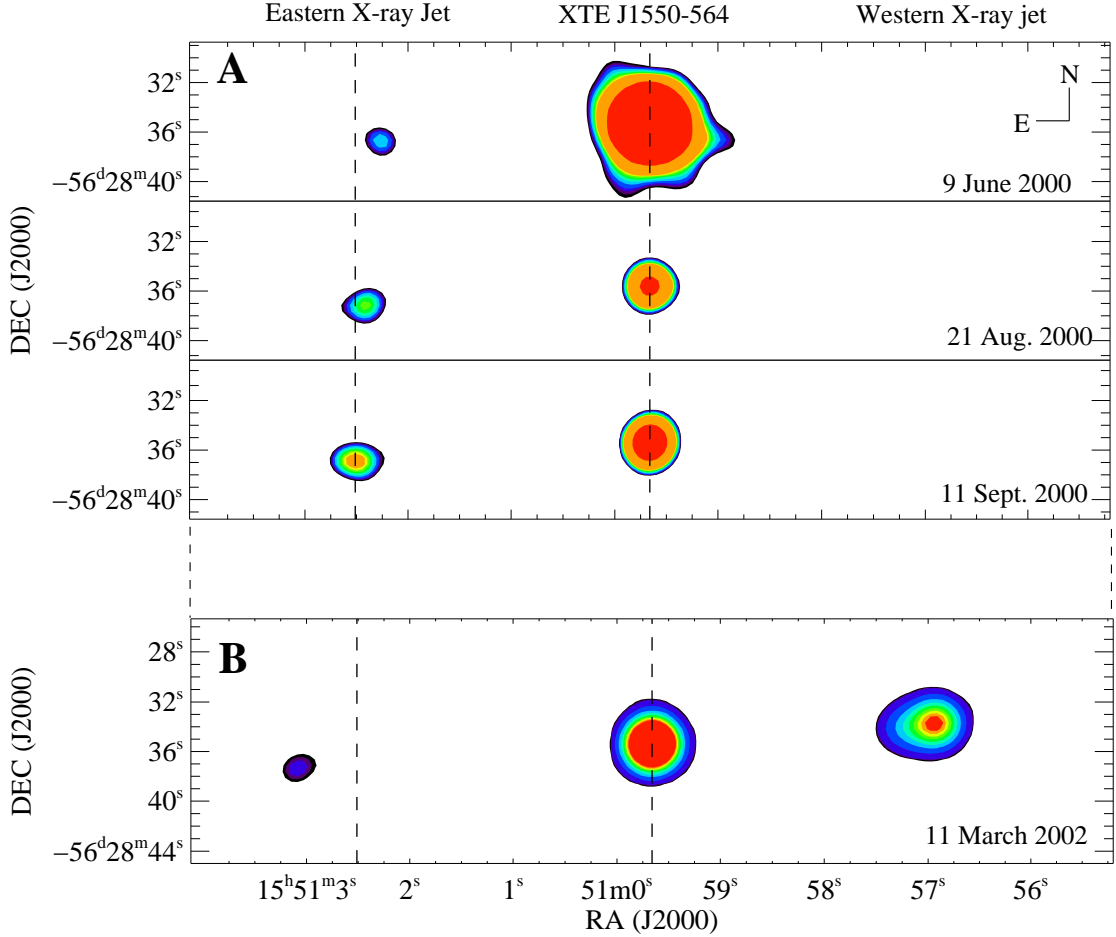}}} \par}  \caption{
{\em Chandra} 0.3--8~keV images (using the Advanced CCD
Imaging Spectrometer spectroscopy array ACIS-S), which show
the black hole binary XTE J1550--564 (center), the
approaching eastern x-ray jet (left) and the receding western
x-ray jet (right).  The observations are ordered
chronologically from top to bottom, and each image is
labeled with the observation date.  These filled contour
plots have been produced by convolving the original Chandra
image with a 2-dimensional Gaussian with a width of 2 pixels
in both directions.  In 2000 ({\bf A}), there are 11 contours
between the lowest contour of $1.33\times 10^{-3}$ count
s$^{-1}$ pixel$^{-1}$ and the highest contour of $8.16\times 10^{-3}$
count s$^{-1}$ pixel$^{-1}$ .  The same contour levels are used in all
three 2000 images, but it should be noted that the flux
levels for 9 June 2000 are not directly comparable to those
for the other two observations because a grating was inserted
for the June 9 observation. For the 2002 image ({\bf B}),
there are 11 contours between $0.33\times 10^{-3}$ count
s$^{-1}$ pixel$^{-1}$ and $8.16\times 10^{-3}$ count s$^{-1}$  pixel$^{-1}$.  The dashed
lines mark the positions of XTE J1550-564 and the eastern
x-ray jet on September 11 when the sources were separated by
23$^{\prime\prime}.4$.  The proper motion of the x-ray jet is
$21.2\pm 7.2$ mas day$^{-1}$ between 9 June 2000 and 11
September 2000  and averages $10.4\pm 0.9$ mas day$^{-1}$
between 11 September 2000 and 11 March 2002, indicating
deceleration of the jet.   Assuming a power-law spectral
shape with a photon  index of 1.7 and interstellar absorption
of N$_\mathrm{H}$ = $9\times 10^{21}$ cm$^{-2}$, a count rate
of $1.33 \times 10^{-3}$ count s$^{-1}$ corresponds to a flux
of $7.71\times 10^{-14}$ erg cm$^{-2}$ s$^{-1}$ in the
0.3-8~keV band for the 9 June 200 observation (with the
grating) and to $1.66\times 10^{-14}$ erg cm$^{-2}$ s$^{-1}$ for 
other observations.}\end{figure*}

\begin{figure*}[p] {\par\centering
\resizebox*{1\textwidth}{!}{
\rotatebox{-0}{\includegraphics{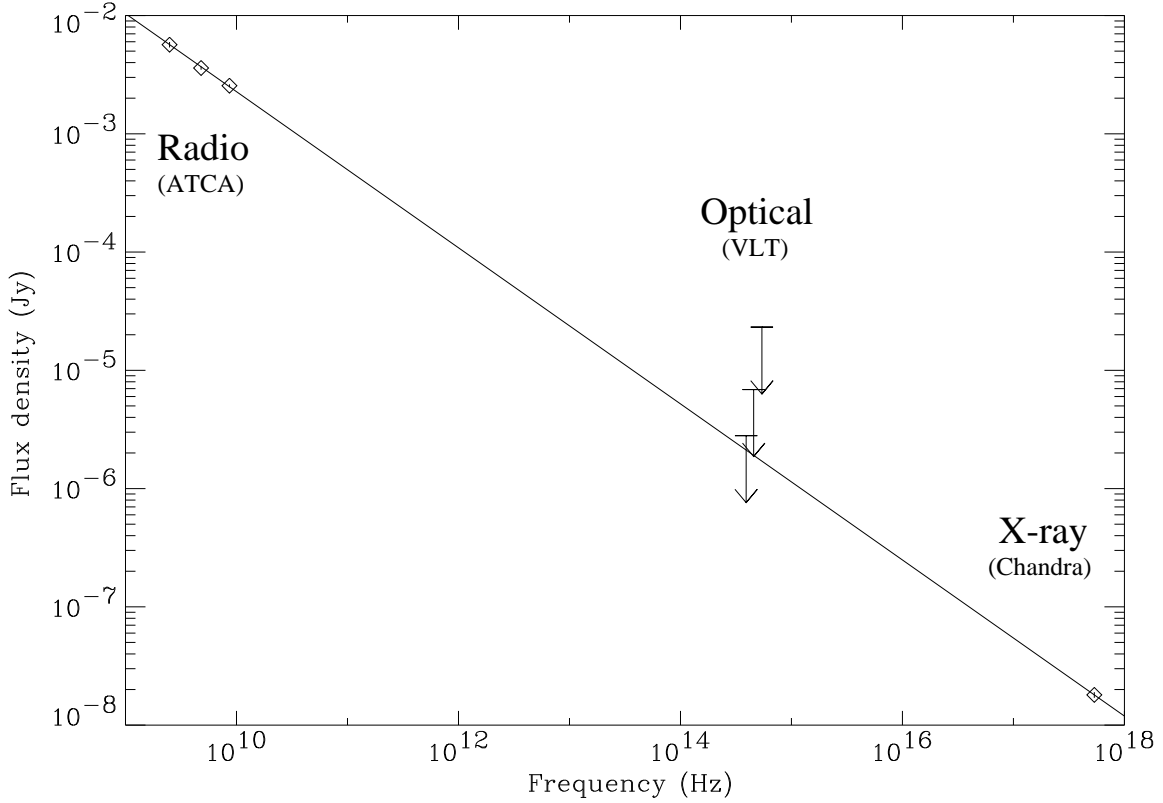}}} \par} \caption{The
spectral energy distribution of the western jet around 2002
March 11.  The radio points near 10$^{10}$ Hz are ACTA
measurements from March 6.  The radio flux density  were 5.7
$\pm$  0.3, 3.60 $\pm$ 0.08 and 2.55 $\pm$ 0.07 mJy at 2496,
4800, 8640 MHz respectively, giving a spectral index of
--0.63 $\pm$ 0.05 in the radio range. The x-ray measurement
near 5 $\times$ 10$^{17}$ Hz is the Chandra measurement from
March 11.  X-ray spectral fitting of the Chandra data for the
western source  assuming a powerlaw form with interstellar
absorption fixed to  the total Galactic HI column density
($N_{\mathrm{H}} = 9.0 \times 10^{21}$ cm$^{-2}$) gives a
spectral index of --0.70 $\pm$ 0.15 (90 \% confidence level).
The spectrum may be somewhat steeper if there is additional
absorption near the source. The unabsorbed 0.3--8 keV flux is
3.8 $\times$ 10$^{-13}$ ergs cm$^{-2}$ s$^{-1}$ (i.e.\ 18~nJy
at 2.2~keV). The optical upper limits (99\% confidence level)
in between are derived from deep observations carried out
with the 8.2 metre Unit 3 telescope at the European Southern
Observatory, Paranal.  The source was observed with the FORS1
instrument in the Bessel $V$ and $R$ filters on March 10,
with limiting magnitudes for point sources of 25.2 and 25.5
mag., respectively, and on March 18 with FORS1 and the Bessel
$I$ filter, with a limiting magnitude for point sources of
25.5 mag. We assumed an optical extinction of $A_V$ = 4.75
mag. \cite{oro02}. The broadband spectral energy distribution
is consistent with a single powerlaw of spectral index
--0.660 $\pm$ 0.005. } \end{figure*}

\end{document}